\documentclass[10pt]{article}
\usepackage{amssymb,amsmath}
\usepackage{graphicx}
\usepackage[margin=1.0in]{geometry}
\usepackage{hyperref}

\def\dsl{\displaystyle}
\title{N-flation in Supergravity}

\author {\sf Kumar Das,\footnote{kumar.das@saha.ac.in} \hspace{4pt} Koushik Dutta,\footnote{koushik.dutta@saha.ac.in} \\[10pt]
\small\em Theory Physics Division, Saha Institute of Nuclear
        Physics, 1/AF Bidhan Nagar, Kolkata 700064, India 
}

\date{}

\begin{document}

    \maketitle
    \begin{abstract}
We have constructed a large field N-flation model in the supergravity framework. In this simple set-up, $N$ fields collectively drive inflation where each field traverses sub-Planckian field values. This has been realised with a generalisation of the single field chaotic inflation in Supergravity. Interestingly, despite of the presence of the field interactions, the dynamics can be described in terms of an effective single field. The observable predictions of our model \emph{i.e.} tensor to scalar ratio $r$ and scalar spectral index $n_s$ are akin to the chaotic inflation. 
\end{abstract}

\vspace*{0.07cm}

\section{Introduction}
Inflation is a paradigm in cosmology that has received much of its support from current observations \cite{Mukhanov:2005sc}. It solves few problems of standard big bang cosmology, and is able to give an explanation of the structure formation in the Universe.  The scalar perturbations responsible for the large scale structures of the Universe are generated due to the quantum fluctuations of the inflaton field,  and the tensor perturbations originate from the fluctuations of the spin-2 graviton field. The tensor amplitude induces $B$-mode polarisation in the CMB temperature anisotropy which is considered to be an unique signature of inflation. In this year, the BICEP II team has announced the first `tentative' detection of primordial B-mode polarisation corresponding to a tensor to scalar ratio $r \sim 0.15$ \cite{Ade:2014xna}.  

Assuming that the source of the $B$-mode polarisation is primordial, the detected value of $r$ points to a scale of inflation that is close to the GUT scale of around $10^{16}$ GeV. Following Lyth bound, it also indicates super-Planckian field excursion ($\Delta \phi > M_{Pl}$ ) during inflation \cite{Lyth:1996im}. First of all, having super-Planckian field range with a cut-off scale of Planck mass is difficult to accommodate in the usual notion of effective quantum field theory. Secondly, it is turning out very difficult, if not impossible to arrange large field range in the ultraviolet complete theory like string theory. Many notable attempts have been made though \cite{McAllister:2008hb}, \cite{Silverstein:2008sg}, \cite{Kaloper:2008fb}, \cite{Kaloper:2011jz}, \cite{McAllister:2014mpa}, \cite{Arends:2014qca}.

Another complimentary approach is to find special direction in the multi-dimensional field space where a particular combination of the field directions is flat enough to allow super-Planckian vev. In the case of two field case, this has been realised in `aligned inflation' where two axion fields have been used \cite{Kim:2004rp}, \cite{Kappl:2014lra}, \cite{Ben-Dayan:2014lca},  \cite{Czerny:2014qqa}, \cite{Li:2014xna}, \cite{Gao:2014uha}. In the case of N-flation, many fields contribute to drive inflation where each field moves over sub-Planckian vev \cite{Dimopoulos:2005ac}. Recently, attempts have been made in incorporating N-flation in the string theory set-up \cite{Kallosh:2007cc}, \cite{Grimm:2007hs}, \cite{Cicoli:2014sva}.  
 
In this work, we propose a model of N-flation in the supergravity (SUGRA) frame work, where each field has a quadratic mass term potential of chaotic inflation. This is a simple generalisation of single field chaotic inflation scenario in SUGRA \cite{Kawasaki:2000yn}. Considering that the individual field range is sub-Planckian, the effective description in SUGRA is well under control as long as we demand that the imposed symmetry is not broken by the ultraviolet degrees of freedoms. One important aspect of our construction would be that even though the fields have interactions among themselves, in the cosmological background they collectively behave like single degree of freedom without any interactions. This is true only because of the particular nature of the interactions dictated by the proposed form of the model. 

After briefly discussing the basic points of inflation in SUGRA, and N-flation in the next two sections, we work out our proposal for two-field case in Section (\ref{nfls}). N-field generalisation would be done in section (\ref{nfls}), followed by the conclusion and discussion at the end. 

\section{Chaotic Inflation in SUGRA}
\label{chaotic}
In this section we will discuss inflationary scalar potential in the SUGRA framework, and as an example, we will outline how chaotic inflation can be realised in SUGRA \cite{Kawasaki:2000yn}. Our N-flation construction is crucially dependent on this elegant proposal, and in fact it is a simple generalisation to $N$ fields.

Being very simplistic in the form of a potential with a mass term, chaotic inflation is an attractive model amongst the zoo of inflationary models. It has also gained some tentative observational support after the release of BICEP II data that hints towards a tensor to scalar ratio $r\sim 0.1$ \cite{Ade:2014xna}. Using Lyth bound \cite{Lyth:1996im} 
\begin{align}
\Delta\phi=\mathcal{O}(1)\times\left(\frac{r}{0.01}\right)^{1/2},
\end{align}
the data immediately requires super-Planckian field excursion during inflation. Now, the super-Planckian field excursion is a natural requirement for chaotic inflationary potential $V(\phi)=\frac{1}{2}m^2 \phi^2$ for slow-roll parameters being small. Thus embedding chaotic inflation in any particle physics set-up is of paramount importance. Here we briefly review how chaotic inflation potential emerges naturally in SUGRA. 

A model in $\mathcal{N}=1$ SUGRA is obtained by specifying two quantities, K$\ddot{\text a}$hler potential $K(\Phi,\bar{\Phi})$ and superpotential $W(\Phi)$, where $W$ is a holomorphic function of fields $\Phi$, and $K$ is a real function \cite{Baumann:2014nda}. Inflaton $\phi$ is a member of the complex chiral superfield $\Phi$, and for simplicity we are assuming inflaton to be singlet under relevant gauge group. This allows us not to worry about $D$-term contribution to the scalar potential. The $F$-term part of the scalar potential is 
\begin{align}
V=e^{K} \left(K^{i\bar{j}} D_i W D_{\bar{j}}\bar{W} - 3 |W|^2 \right),
\label{scl}
\end{align}
where $K^{i\bar{j}}$ is the inverse of the K$\ddot{\text a}$hler metric $K_{i\bar{j}}=\frac{\partial^2 K}{\dsl{\partial\Phi_i \partial\bar{\Phi}_{\bar j}}}$ and $D_i W=\frac{\partial W}{\partial\Phi_i} + W\frac{\partial K}{\partial\Phi_i}$.
 The kinetic term in the Lagrangian of the scalar field is 
\begin{align}
\mathcal{L}_{K.E.}=K^{i\bar j}(\partial_{\mu}\Phi_i)(\partial^{\mu}\bar{\Phi}_{\bar j}).
\end{align}
Here we worked in Planck units where the reduced Planck mass $M_{Pl}\simeq 2.4\times 10^{18}$ GeV is set to be unity. 
For the cannonical choice of $K=\Phi\bar{\Phi}$, the F-term potential $V\propto e^{|\Phi|^2}$, and its slope is too steep to sustain the flatness required for chaotic inflation. This is the so called $\eta$-problem in F-term inflation \cite{Copeland:1994vg}, \cite{Lyth:1998xn}. The elegant solution was proposed in \cite{Kawasaki:2000yn} where the authors introduced a shift symmetry to the K\"ahler potential of the complex chiral superfield $\Phi$, under which $\Phi \rightarrow \Phi+iC,$ where $C$ is some real constant. This restricts the form of $K$ to be a function of Re$(\Phi)$ only. In this case, Im$(\Phi)$ does not appear in the exponential, and it can be identified as inflaton free of $\eta$-problem\footnote{Heisneberg symmetry can also be imposed in solving $\eta$-problem \cite{Gaillard:1995az, Antusch:2008pn, Antusch:2009ty}.}.  

In \cite{Kawasaki:2000yn} the following superpotential and K\"ahler potential was proposed
\begin{align} 
W=mX\Phi, ~~~~K=X\bar{X}-\zeta(X\bar X)^2 - \frac 1 2 (\Phi\pm\bar{\Phi})^2, \label{chaotic_sugra}
\end{align}
where $\Phi$ contains the inflaton field and $X$ is an auxiliary chiral superfield that remains at zero vev during inflation. The superpotential breaks the shift symmetry imposed in the K\"ahler potential, and thus gives rise to the tree-level mass term via the $F$-term of the auxiliary field $X$. In other words supersymmetry is broken along the $X$-direction ($D_{X}W\neq 0$) and it supplies the necessary potential energy to drive inflation. Here $-\zeta(X\bar X)^2$ term is added to render the mass of the  $X$ field being greater than the Hubble scale. For $\zeta=0$, the mass of the $X$ field is comparable to the mass of the inflaton. So there will be inflationary fluctuations of $X$ during inflation and hence the dynamics can not be regulated with one field. The above construction is characterised by $W_{inf} = 0, D_{X}W_{inf}\neq 0,  D_{\Phi}W = 0$ during inflation, and it has been generalised in the case of hybrid inflation scenario in the framework of tribrid inflation \cite{Antusch:2009vg, Antusch:2009ef, Kallosh:2010xz}

From the point of effective SUGRA theory, this set-up is complete in a sense that with the assumption of shift symmetry breaking term in the superpotential is small, its corrections (potentially shift symmetry breaking) to the K\"ahler potential are going to be also parametrically small. Now the smallness of the symmetry breaking parameter $m$ is ensured by the scalar amplitude of density fluctuations. This fixes the value of $m \sim 10^{-5} M_{pl}$. Whether the symmetry breaking is under control in any UV complete theory like string theory is an open issue, and it requires understanding of the dynamics of stringy degrees of freedom. 

Even though the construction is elegant, the difficulties behind the description of large field inflation with one single field is problematic from the point of view of purely effective field theories. In the context of effective field theory, the inflation potential can be written in the following form
\begin{align}
V_{eff}=V(\phi)+\sum_{n = 0}^{\infty}c_{n}V(\phi)\frac{\phi^{n + 1}}{M_{Pl}^{n + 1}},
\end{align}
where $c_n$'s are dimensionless coefficients of order one. Since the inflaton has to traverse over a trans-Planckian distance in the field space during the time inflation takes place, \emph{i.e.} $\Delta\phi>M_{pl}$, each term in the summation contributes equally well to the potential unless its coefficients $c_n$'s are finely tuned. So large field inflation becomes sensitive to an infinite number of such terms. If we want to predict the dynamics, we necessarily need to know all these terms. Note that in the setup of Eq.~(\ref{chaotic_sugra}), these higher dimensional operators are under control due to the imposed symmetry and its soft breaking in the superpotential. In the next section, we will discuss how the problem of super-Planckian vev of a single field can be evaded in the set-up of N-flation by distributing the job of driving inflation in $N$ fields \cite{Dimopoulos:2005ac}. 

\section{N-flation}
\label{nfl}
Even though the formulation of chaotic inflation in SUGRA is well understood, the tentative observations require the field associated with the single inflaton to be super-Planckian. Thus constructing a model of inflation with more than one field is worth formulating where the job of driving inflation is distributed among many $\phi_i$ fields. Each $\phi_i$ satisfies $\Delta\phi_i<M_{Pl}$ 
Under this condition the potential for an individual field can be expanded in the effective field theory framework. As we will see, in the proposed N-flation scenario, the total field displacement is now
\begin{equation} 
 \Delta\phi_{total}^2=\sum_{i=1}^{N}\Delta\phi_i^2 > M_{Pl}.
\end{equation}

The actual idea of N-flation was first proposed by Dimopoulos et.al \cite{Dimopoulos:2005ac}, and the basic point was inflaton is a collection of $N$ number of fields that drives inflation through assisted inflation mechanism \cite{Liddle:1998jc, Jokinen:2004bp, Copeland:1999cs}, rather than a single field. Here each field $\phi$ has a field excursion smaller than Planck scale. Individual fields are not capable of producing the slow-roll for an appreciable number of e-folds, but a collection of such fields produces sufficient e-folds to solve the cosmological problem. So the dynamics is determined collectively by $N$ such fields. The motivation was from the standpoint of particle physics where the existence of scalar field is ubiquitous. In the original work of \cite{Dimopoulos:2005ac}, the individual field was axion having periodic potential. Around the bottom of the potential where inflation happens, the potential was written as the sum of potential for each individual field \emph{i.e.}
\begin{align}
V(\phi_i)=\sum_{i=1}^{N}V_{i}(\phi_i)=\sum_{i=1}^{N}\frac 1 2 m^2 \phi_{i}^2
\end{align}
So here each field $\phi_i$ is moving under the potential $m^2\phi_{i}^2 /2$. It was assumed that the cross-couplings between the fields are negligible. Considering an initial configuration where each field is displaced from the minimum of the potential by a sub-planckian displacement $\langle\phi_{n0}\rangle=\alpha_{n}M_{Pl}$, the total displacement in field space in polar coordinate is 
\begin{align}
\rho^2 =\sum_{i}\phi_{i}^2 =\sqrt{N}\alpha M_{Pl}.
\end{align}  
In term of the variable $\rho$ the effective lagrangian density is,
\begin{align}
\mathcal{L}\simeq (\partial\rho)^2 + \rho^2 (\partial\Omega)^2 - \frac 1 2 m^2 \rho^2.
\end{align}
Now the the angular degree of freedom $\Omega$ has no potential energy, and its equation of state parameter $\omega_{\Omega}=1$. Thus its energy density falls as $a^{-6}$, and its contribution becomes negligible soon compared to the radial field $\rho$. Effectively, the radial variable $\rho$ will act as an inflaton with its single field dynamics.

Our objective is to construct a model of N-flation in the framework of SUGRA. In our simple frame-work, the fields are going to have cross couplings among themselves. But because of the particular nature of the coupling that automatically arises in the set-up, the effective field dynamics in the cosmological background is similar to the single field. In our case the potential will be like
\begin{align}
\label{form}
V(\phi_1,...,\phi_N)=\sum_{i=1}^{N}V_{i}(\phi_i) + interactions.
\end{align}
We will first discuss a simple case where the inflaton is a collection of a pair of fields. In this case, we will solve the dynamics numerically to show that the field trajectory in the slow-roll attractor is a straight line. Then we will go for the generalisation with $N$ fields.

\section{N-flation in SUGRA}
\label{nfls}
We present our main result in this section. As a toy example we first analyse the two-field case, where trivial redefination of fields can make the dynamics effectively single field. We also analyse the background dynamics numerically to show how the attractor solution emerges for the effective single field case. This shows that the dynamics is governed by one degree of freedom. Subsequently, we do the generalisation for N-fields, where we analytically prove how it can be reduced to the single field.   

\subsection{Two field Case}
Let us begin with the following choice of the superpotential
\begin{align}
W=mX(\Phi_1 + \Phi_2) \label{superpotential}
\end{align}
Here each $\Phi$ is a chiral superfield which contains one singlet inflaton field. The masses of the fields are taken to be degenerate for simplicity. $X$ is another chiral superfield that is needed to provide the vacuum energy via its non-zero F-term. The K$\ddot{\text{a}}$ler potential is taken to be
\begin{align}
K= X\overline{X}-\zeta(X\overline{X})^2 - \frac 1 2 (\Phi_1 - \overline{\Phi}_1)^2 - \frac 1 2 (\Phi_2 - \overline{\Phi}_2)^2   
\end{align}
Here $-\zeta(X\overline{X})^2$ term is added for the stabilization of $X$ field as mentioned in detail in section (\ref{chaotic}). This will ensure that the $X$ field gains a mass larger than the inflaton mass during inflation and hence it will not disturb the inflationary dynamics. The K$\ddot{\text{a}}$hler potential respects the shift symmetry for the inflaton fields: $\Phi_i \rightarrow \Phi_i + i \alpha_i$. The real components of those fields can be identified as inflaton fields. This is to avoid the usual $\eta$-problem. 

Let us now decompose the complex superfields $\Phi_1$ and $\Phi_2$ into a pair of real scalar fields
\begin{equation}
\Phi_1=\frac 1 {\sqrt 2}(\phi+i\beta),~~
\Phi_2=\frac 1 {\sqrt 2}(\chi+i\sigma).
\end{equation}  
The masses for these fields can be calculated using the F-term SUGRA expression given in Eq.~(\ref{scl}).
The masses-squared of the field $X$ (both real and imaginary parts) is given by 
\begin{align}
m_{X}^2=12\zeta H^2 + 2m^2
\end{align} 
So for positive $\zeta\neq 0$, $m_X > \mathcal{O}(H)$, and it decouples from the inflationary dynamics in settling to its minima at $X=0$.  Now in the trajectory of $X=0$, the mass squared of the fields $\text{Im}~\Phi_1$ and $\text{Im}~\Phi_2$ are
\begin{equation}
m_{\beta}^2 =m^2 (1+2\sigma^2)+6H^2, ~~~m_{\sigma}^2 =m^2(1+2\beta^2)+6H^2. 
\end{equation} 
Therefore during inflation they are also stabilized at $\beta=\sigma=0$. We have checked the stabilisation numerically by solving the dynamics.

There are two inflaton fields $\phi$ and $\chi$ whose potential do not contain the $e^K$ factor of Eq.~(\ref{scl}) and thereby evade the $\eta$-problem. Along the inflationary trajectory $X=\beta=\sigma=0$, the scalar potential as computed from Eq.(\ref{scl}) looks like
\begin{align}
V(\phi,\chi)=\frac 1 2 m^2 \phi^2 +\frac 1 2 m^2 \chi^2 + m^2 \phi\chi. \label{pot}
\end{align}
Clearly, this potential is a sum of two chaotic inflation potential together with an interaction term as written earlier in Eq.~(\ref{form}). At this point, we draw particular attention to the nature of the coupling which depends on each field linearly. Including the kinetic terms, the Lagrangian density for the inflaton fields is given by
\begin{align}
\mathcal{L}=(\partial_{\mu}\phi)(\partial^{\mu}\phi) + (\partial_{\mu}\chi)(\partial^{\mu}\chi) - V(\phi,\chi),
\end{align}
where $V(\phi,\chi)$ is given by Eq.\eqref{pot}.

The above Lagrangian can be easily casted in a more convenient form by defining two new fields
\begin{equation}
\varphi_1 = \frac 1 {\sqrt{2}}(\phi+\chi), ~~
\varphi_2 = \frac 1 {\sqrt{2}}(\phi-\chi),
\end{equation}
and in terms of these two fields the Lagrangian density can be written as,
\begin{align}
\mathcal{L}=(\partial\varphi_1)^2 + (\partial\varphi_2)^2 - \frac 1 2 (\sqrt 2 m)^2 \varphi_{1}^2.
\end{align}
Here one degree of freedom $\varphi_1$ is massive while the other one $\varphi_2$ is massless. In the two dimensional field space, there is a flat direction $\varphi_2$ along which the potential vanishes. So the equation of state parameter $\omega =1$ for $\varphi_2$. Thus its associated energy density redshifts very quickly and becomes cosmologically irrelevant, 
\begin{align}
\dsl{\rho_{\varphi_2}}\propto a^{-3(1+\omega_{\varphi_2})}\propto a^{-6}.
\end{align}
On the other hand $\varphi_1$ direction has a chaotic inflation potential that can drive inflation. The total field variation $\Delta \varphi_1 \sim \Delta \phi+\Delta \chi$. It is important to note that this is crucially different from the original N-flation set-up, where for two field case the effective displacement of the radial field direction is always positive $\Delta \varphi_1^2 \sim \Delta \phi^2+\Delta \chi^2$ \cite{Dimopoulos:2005ac}. In our case, this is not necessarily true, and we consider this point as one drawback of our set-up. In our setup, the effective displacement of the field is dependent on the initial field configurations. At this point, we note that reduction of the potential with coupling term into an effective single-field case is possible only for the $m^2\phi\chi$ coupling. This is no longer true for $\phi^2\chi^2$ coupling where it is evident that field trajectory is curved in general, and can not be described by one degree of freedom \cite{Peterson:2010np}.

\begin{figure}[h!]
\centering
\includegraphics[width=72mm]{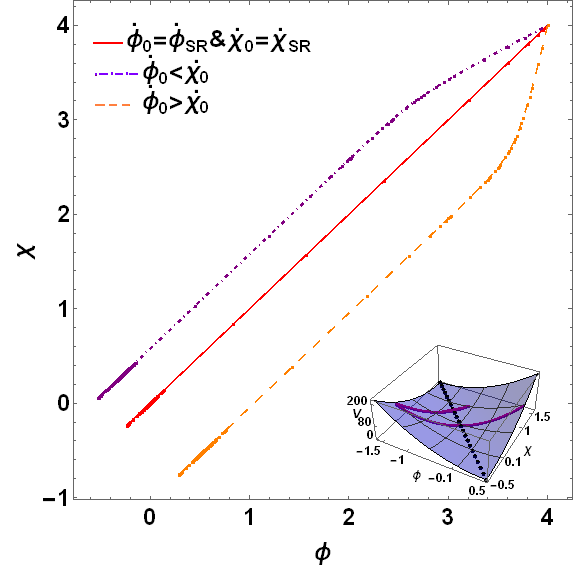}
\caption{Field space plots for three different boundary conditions. In the inset, we show the evolution of the field for single trajectory near its minima. (For interpretation of the references to color in this figure legend, the reader is referred to the web version of this article.)}
\label{combplot}
\end{figure}

\begin{figure}[h!]
\centering
\includegraphics[width=82mm]{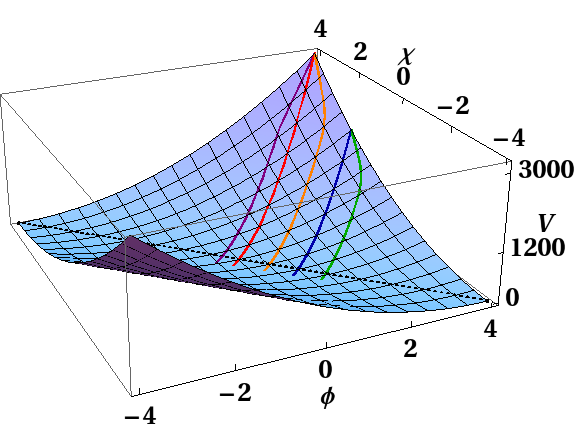}
\caption{Effective potential showing the joint evolution of fields $\phi$ and $\chi$. (For interpretation of the references to color in this figure legend, the reader is referred to the web version of this article.) }
\label{potnplot}
\end{figure}

In Fig~(\ref{combplot}), we show the field space plot of this simple set-up. We have analysed three cases that corresponds to three different initial boundary conditions to elucidate the nature of the attractor behaviour in this case. As we see the attractor solutions in the field space are straight lines. Here each field satisfies the same Friedmann equation. During  slow roll of $\phi$ and $\chi$ the field displacements for $\phi$ will be proportional to the field displacement for $\chi$ in the phase space \emph{i.e.} $\Delta\phi\propto\Delta\chi$ in same time interval. So $\phi=\chi+c$ will be the attractor solution, where $c$ is a constant. In the field space this will be reflected in straight line behaviour which is what we obtained in Fig.(\ref{combplot}). 

Now the potential can be written as $V(\phi,\chi)=\frac 1 2 m^2 (\phi+\chi)^2$. The minima of this potential is not a single point but a line whose equation is $\phi+\chi=0$. This equation is satisfied by many field points $(\phi,\chi)$. The central plot (red) in Fig.(\ref{combplot}) corresponds to the initial conditions $\phi_0 =\chi_0 =4M_{pl}$ and $\dot{\phi}_0 =\dot{\phi}_{SR}$ and $\dot{\chi}_0 =\dot{\chi}_{SR}$. In this case, equation of minima is $2\phi=0$ or $2\chi=0$. So minimum of $V(\phi,\chi)$ is at the origin. That is why the central plot passes through the origin. Now we keep the former initial condition intact and change the later to $\dot{\phi}_0 \neq\dot{\phi}_{SR}$ and $\dot{\chi}_0 \neq\dot{\chi}_{SR}$. Then two cases may appear one is $\dot{\phi}_0 <\dot{\chi}_0$ (purple) and the other is $\dot{\phi}_0 >\dot{\chi}_0$ (orange). For the first case (shown in purple) the trajectory is initially curved and then parallel to the central straight line (red). The reason for this is that as we have started with a different $\dot{\phi}$ so it will first meet the attractor when it slow rolls and then its subsequent evolution in the phase space is similar to the case of a single-field chaotic inflation. So is the case with $\dot{\chi}$. But as at $t=0$ we have given different initial values to $\dot{\phi}$ and $\dot{\chi}$ the manner in which both $\dot{\phi}$ and $\dot{\chi}$ converge to their respective attractor solution is not the same. This is manifested in the curvature of the purple coloured plot. The explanation is same for the second case $\dot{\phi}_0 >\dot{\chi}_0$ (orange) too as the initial condition is just the opposite of the first. Near the lower left corner of Fig.(\ref{combplot})there is a little thick portion in each of these lines. This is where the fields oscillate around the minima of $V$. The minimum is different for three lines as explained earlier. The behaviour of the fields near the minima of the potential is shown in the small box at the lower right coner in this figure. This represents the oscillatory part of the field evolution.  
  
The Fig.~(\ref{potnplot}) is a 3D plot that shows the combined evolution of the fields $\phi$ and $\chi$ in the effective potential $V(\phi,\chi)$. The field trajcetories of Fig.(\ref{combplot}) is shown on the potential. Basically Fig.(\ref{combplot}) is a two dimensional projection of
Fig.(\ref{potnplot}) in $\phi-\chi$ plane. The dotted black line represents the minima of the potential. Two additional field trajectories are shown in Fig.(\ref{potnplot}) for $\phi_0 >\chi_0$. Blue line corresponds to the initial condition $\dot{\phi}_0 =\dot{\phi}_{SR}$ and $\dot{\chi}_0 =\dot{\chi}_{SR}$ and green line corresponds to the condition $\dot{\phi}_0 >\dot{\chi}_0$. All these family of straight lines depicts the attractor behaviour. In summary, we note that $\phi = \chi + constant$ are attractor solutions, and even if the field has non-slow-roll initial conditions, it reaches to the attractor solution quickly and follow the straight line trajectory in showing its effective one degree of freedom.


\subsection{N-Field Generalization}
Now we are going to extend our formalism for $N$-field configuration in $\mathcal{N}=1$ SUGRA. Here the involvement of $N$-fields share among themselves the complete task of producing super-Planckian field excursion during inflation. We propose the following superpotential
\begin{align}
W=mX\sum_{i=1}^N \Phi_i , \label{superpotential1}
\end{align}
where $\{\Phi_i \}$'s are the set of complex chiral superfields and $i\in\mathbb I$. Each $\Phi_i$ contributes one inflaton. The K$\ddot{\text{a}}$hler potential is 
\begin{align}
K=(X\bar X)-\zeta(X\bar X )^2 - \frac 1 2 \sum_{i=1}^N (\Phi_i - \bar{\Phi}_i )^2
\end{align}
Shift symmetry is satisfied by every chiral field \emph{i.e.} $\Phi_i \to\Phi_i + iC,\,\,\,\forall\,i$, $C$ is real constant. Now each complex $\Phi_i$ can be decomposed into a real and imaginary parts like two-field case. Then along the diretion of inflation $X= Im~\Phi_1=Im~\Phi_2=...=Im~\Phi_N=0$. The scalar potential in this case takes the form
\begin{align} 
V&=\frac 1 2 m^2 \left(\sum_{i=1}^N \phi_i \right)^2 \\ 
&=\frac 1 2 m^2 \sum_{i=1}^N \phi_{i}^2 + \text{two field interaction},
\end{align}
where $\phi_i=Re~\Phi_i,\,\forall\,i.$ Masses of the fields are degenerate. Now we will find an orthogonal combination of $\Phi_i$'s such that the potential becomes a single-field potential.

Let us define a vector,
\begin{equation}
\bf{\Phi}=\left( 
\begin{array}{c}
\phi_1 \\ \phi_2 \\ \vdots \\ \phi_N 
\end{array}
\right)
\end{equation}
Here the set of basis vectors $\bf{\{e_i\}}$ are 
\begin{equation*}
|e_1 \rangle=
\begin{pmatrix} 
1\\0\\0\\ \vdots\\0
\end{pmatrix},\quad
|e_2 \rangle=
\begin{pmatrix} 
0\\1\\0\\ \vdots\\0
\end{pmatrix},
\end{equation*}
\begin{equation*}  
|e_3 \rangle=
\begin{pmatrix} 
0\\0\\1\\\vdots\\0
\end{pmatrix},\quad\hdots\quad,
|e_N \rangle=
\begin{pmatrix} 
0\\0\\0\\ \vdots\\1
\end{pmatrix}
\end{equation*}
Let us move to a different set of basis vectors $\bf{\{e_i'\}}$. With respect to this new basis vector the components of $\bf{\Phi}$ will also change accordingly. Thus we can write,
\begin{align}
\bf{\Phi}=\phi_i |e_i \rangle 
=\psi_j |e_j' \rangle \label{we}
\end{align}
The old basis is related to new basis by 
\begin{align}
|e_i' \rangle &=\bf{T_{ij}} |e_j \rangle \label{new} \\
\bf{T_{ij}} &=\dsl{\langle e_j |e_i' \rangle} \quad \text{so}\quad \bf{\Phi}=\underbrace{\phi_i T_{ij}^{-1}} |e_j' \rangle \label{are}
\end{align}
Comparing eqns.(\ref{we}) and (\ref{are}) we find
\begin{align}
\psi_j=\phi_i \bf{T_{ij}^{-1}} \label{com}
\end{align}
As we aim at reducing the $N$-field Lagrangian density into a single field Lagrangian density so we define,
\begin{align}
\psi_1 =\frac{1}{\sqrt{N}} (\phi_1 +\phi_2 + \hdots +\phi_N)
\end{align}
From eqn.(\ref{com}) 
\begin{align}
&\psi_1 = T_{11}^{-1}\phi_1 + T_{21}^{-1}\phi_2 + \hdots + T_{N1}^{-1}\phi_N \\
\text{So,} \qquad &T_{11}^{-1}=T_{21}^{-1}=T_{31}^{-1}\hdots\hdots=T_{N1}^{-1}=\frac{1}{\sqrt N}
\end{align}
But as $T_{ij}$ is an orthogonal matrix \emph{i.e.} $T_{ij}^{-1}=T_{ji}=T_{ij}^{T}$, so, $T_{11}=T_{12}=\hdots=T_{1N}=\frac{1}{\sqrt N}$. Now we obtain from eqn.(\ref{new})
\begin{equation}
|e_1' \rangle =
\frac{1}{\sqrt N}(|e_1 \rangle +|e_2 \rangle +\hdots+ e_N \rangle)=
\frac{1}{\sqrt N}\begin{pmatrix} 1\\1\\\vdots\\1\end{pmatrix} 
\end{equation}
Now by Gram-Schmidt process the other orthonormal basis vectors can be found out and they are,
\begin{eqnarray*}
|e_2' \rangle&=&\frac{1}{\sqrt{\langle e_2'|e_2' \rangle}}\left(|e_2 \rangle - \frac{\langle e_1'|e_2 \rangle}{\langle e_1'|e_1' \rangle}\,|e_1' \rangle\right) \\
|e_3' \rangle&=&\frac{1}{\sqrt{\langle e_3'|e_3' \rangle}}\left(|e_3 \rangle - \frac{\langle e_1'|e_3 \rangle}{\langle e_1'|e_1' \rangle}\,|e_1' \rangle - \frac{\langle e_2'|e_3 \rangle}{\langle e_2'|e_2' \rangle}\,|e_2' \rangle\right) 
\end{eqnarray*}
\\
$~~~\qquad\hdots \qquad \hdots \qquad \hdots \qquad \hdots \qquad \hdots$
\\
In general,
\begin{equation*}
|e_N' \rangle=\frac{1}{\sqrt{\langle e_N'|e_N' \rangle}}\left(|e_N \rangle - \sum_{i=1}^{N-1}\frac{\langle e_i'|e_N \rangle}{\langle e_i'|e_i' \rangle}\,|e_i' \rangle\right)
\end{equation*}
With respect to these new basis $\bf{\{e_i'\}}$, the new components $\psi_i$'s can be found out using eqns.(\ref{com}) and (\ref{new}). The Lagrangian density is therefore,
\begin{align}
\mathcal{L}=\sum_{i=1}^N (\partial\psi_i )^2 - \frac 1 2 (\sqrt N m)^2 {\psi_1 }^2 \label{lst}
\end{align} 
Thus our method also worked successfully for $N$-field configuration. Here we see from eqn.(\ref{lst}) all fields $\psi_i$, where $i\in[2,N]$, other than $\psi_1$ has no potential energy term. So, they will be easily overthrown from the dynamics since their energy density falls faster than that of $\psi_1$. So, again
\begin{align} 
\mathcal{L}_{eff}=(\partial\psi_1 )^2 - \frac 1 2 (\sqrt N m)^2 {\psi_1 }^2
\end{align}

The interaction term which initially appeared in the expression of the F-term scalar potential does not affect the evolution of $\psi_1$. Only the mass of $\psi_1$ field is enhanced by a factor of $\sqrt N$. Thus the effective dynamics in the attractor solution is governed by one degree of freedom $\psi_1$. Here the total field variation in the field space $\Delta \psi_1 \sim \sum \limits_{i=1}^N \Delta \phi_i$. It is indeed true that the total field variation is dependent on the initial field configuration, i.e initial conditions for the individual field. But in the multidimensional field space, it is expected that the sum can be easily super-Planckian even with some cancellations with sub-Planckian field ranges for individual field. 

\section{Conclusion and Discussions}

With the announcement of tentative observations of large primordial tensor modes by the BICEP II experiments, there is a surge of interest in constructing models of inflation where field excursion is super-Planckian. In the single field set-up, this is problematic from the point of view of effective field theory. Therefore, one option is to use the effects of multiple fields. With tuned parameters, two fields can achieve this large field excursion in the context of natural inflation \cite{Kim:2004rp, Kappl:2014lra, Ben-Dayan:2014lca}. Another possibility is the use of multiple fields having collective dynamics. In the context of many axions in string theory, N-flation was proposed \cite{Dimopoulos:2005ac}. Even though each axion has a periodic potential protected by a perturbative shift symmetry, the inflation happens at the bottom of the potential where the potential can be safely approximated by the chaotic form of quadratic potential. 

In this work, we propose a simple realisation of N-flation in SUGRA. This is based on the generalisation of the set-up of chaotic inflation is SUGRA where $\eta$-problem is solved by the shift symmetries of each individual field. Even though the effective potential has couplings between all fields, but the nature of the couplings allows us to reduce the potential in effective one degree of freedom in the cosmological background. The model has only one free parameter $m$ that controls the breaking of the shift symmetry. Its value is fixed by the normalisation of scalar density perturbations. For the case of two field case, we have solved the background dynamics numerically to show that the attractor behaviour of the solutions, and we have found that the field trajectory is straight line on the attractor solution. For the case of N-fields, reduction to single field has been done analytically. The model has similar predictions to the single-field chaotic inflation case: tensor to scalar ratio $r \sim 0.1$ and $n_s \sim 0.96$. Obviously, the produced density perturbations is of adiabatic type. As we have noted earlier, in our set-up $\Delta \phi_{eff} = \sum_{i}^{N}\Delta \phi_i$, showing that the effective field range in the multi-dimensional field space is dependent on the initial field configurations. We would also like to note that the produced density perturbations would be adiabatic in nature. This is true because the effective dynamics can be well described by one degree of freedom in the attractor solution.

Our set-up involves $N$ number of fields, and therefore it is natural to think that there would be isocurvature modes which are highly constrained by Planck data. It is indeed true that the possibility of existence of isocurvature mode arises only wh
en there are multiple degrees of freedom that carry energy density during inflation and are at the same time lighter than the Hubble constant during inflation. But isocurvature mode can be best understood by analysing the field dynamics in the multi-dimensional field space\cite{Gordon:2000hv}. In this case, the isocurvarure mode can be associated with the curvature of the field trajectory. In other way, if the field trajectory is one parameter family of lines, the only relevant perturbations are adiabatic. This is clear for the simplest two-dimensional case. In our case, as we have shown, the effective field dynamics is one dimensional in the slow-roll trajectory. Other than analytical proof, it has been shown with the numerical analysis for two fields. Therefore, the only source of isocurvature modes is when the field has not reached the attractor solution. Assuming long enough epoch of inflation, this initial phase can be neglected. This argument is similar to the original work of N-flation\cite{Dimopoulos:2005ac} where the perturbations are only of adiabatic type. In summary, we also conclude that in our set-up, only the adiabatic mode is relevant to density perturbations, and for all practical purposes isocurvature perturbations can be neglected.   

At the end of inflation, the potential is going to have many massless degrees of freedom along which the potential is flat. But it is expected that the flat directions are going to be lifted with the soft masses related to low energy SUSY breaking. Explicit nature of these masses can be understood only when N-flation is considered in conjunction with the SUSY breaking sector along the line of \cite{Kallosh:2011qk}. We note that the large number of light species in a given theory typically makes corrections to the Planck mass proportional to $\sqrt{N}$, and this can potentially increase the effective slow-roll parameters spoiling N-flation \cite{Dimopoulos:2005ac}. To answer this question in a concrete manner, we need a complete ultraviolet theory where these corrections can be reliably computed \cite{Cicoli:2014sva}. In the effective SUGRA set-up, this question can not be addressed. Possible resolutions those that are protected from UV physics have been discussed in \cite{Ashoorioon:2009wa, Germani:2010hd}.

In our set-up, we have considered the most simple generalisation of the single field chaotic inflation set-up in SUGRA. Other generalisation of Eq.~(\ref{superpotential1}) are possible. For example, instead of having a common auxiliary field $X$, we may have $X_i$ for each $\Phi_i$, or the mass parameter $m$ can be different. It would be interesting to explore the outcome of those constructions. But in this case, it is expected that the simple analytical understanding as we have done is not possible, and a statistical approach would be more suitable \cite{Easther:2005zr}. 

In summary, we have proposed a simple model of N-flation in SUGRA where chaotic inflation has been realised. In our set-up there are field interactions, but we have shown that the dynamics can be described in terms of an effective single field.

\section{Acknowledgement}
K. Das is supported by the fellowship from the Saha Institute of Nuclear Physics. K. Dutta is partially supported by the Ramanujan Fellowship, and the Max Planck Society-India visiting fellowship. We thank Atanu Kumar, Akhilesh Nautiyal and David Nolde for useful discussions.


\end{document}